\documentclass{article}
\usepackage{dafne99pro}

\usepackage{psfig}

\newcommand{\PRL}[1]{Phys.\ Rev.\ Lett.\ {#1}}

\begin{document}

\title{\bf THE \boldmath $\phi\rightarrow\pi^+\pi^-$ \unboldmath 
AND \boldmath $\phi$ \unboldmath RADIATIVE DECAYS WITHIN A CHIRAL
UNITARY APPROACH}
 
 \author{E. Oset\\
  {\em Departamento de F\'{\i}sica Te\'orica and I.F.I.C.,
  Centro Mixto}\\
  {\em  Universidad de Valencia-C.S.I.C.,
  46100 Burjassot (Valencia), Spain.}\\
 S. Hirenzaki\\
  {\em Department of Physics, Nara Women's University,
  Nara 630-8506, Japan}\\
 E. Marco\\
  {\em Physik Department, Technische Universit\"at M\"unchen,
  D-85747, Germany}\\
 J. A. Oller\\
  {\em Forschungszentrum J\"ulich, Institut f\"ur 
  Kernphysik (Theorie),}\\
  {\em D-52425 J\"ulich, Germany.}\\
 J. R. Pel\'aez\\
  {\em Departamento de F\'{\i}sica Te\'orica,}\\
  {\em Universidad  Complutense de Madrid, 28040 Madrid, Spain}\\
 H. Toki\\
  {\em RCNP, Osaka University, Ibaraki,}\\
  {\em Osaka 567-0047, Japan}}

\maketitle

\baselineskip=11.6pt

\begin{abstract}
 We report on recent results on the decay of the $\phi$ 
 into $\pi^+\pi^-$
  and $\phi$ radiative decays into $\pi^0 \pi^0 \gamma$ and 
   $\pi^0 \eta \gamma$, which require the consideration of the final state
   interaction of a pair of mesons in a region inaccessible to Chiral 
   Perturbation Theory. By using
  nonperturbative chiral unitary methods for the meson meson interaction 
  we can obtain the
  corresponding decay widths and the results are compared with recent
  experimental data.
\end{abstract}
\baselineskip=14pt
\section{The \boldmath $\phi\rightarrow\pi^+\pi^-$ \unboldmath decay}
    The $\phi$ decay into $\pi^+ \pi^-$ is an example of isospin
  violation. The $\phi$ has isospin $I=0$, spin $J=1$, and hence 
it does not couple to the  
  $\pi^+ \pi^-$  system in
  the isospin limit, which implies the rule
  $I+J=\hbox{even}$. The experimental
situation on this decay is rather confusing. There are two
older results whose central values are very
different but their quoted errors
are so big that both were still compatible:
The first   one from \cite{vaser} gives
  $BR=(1.94+1.03-0.81)\times 10^{-4}$. The second one 
  from \cite{golu} provides $BR=(0.63+0.37-0.28)\times 10^{-4}$.
Very recently two new, more precise, 
but conflicting results have been reported
from the two experiments at the VEPP-2M in Novosibirsk:
the CMD-2 Collaboration reports 
a value $BR=(2.20\pm0.25\pm0.20)\times 10^{-4}$ \cite{CMD2}
whereas the SND Collaboration \cite{SND} obtains
$BR=(0.71\pm0.11\pm0.09)\times 10^{-4}$.

Isospin violation has become a fashionable topic in 
Chiral Perturbation Theory ($\chi PT$) \cite{weinberg,xpt} 
but the $\phi\rightarrow\pi\pi$ decay 
is however  unreachable with plain $\chi PT$, since it involves
the propagation of the pair of pions around 1 GeV, far away from 
the $\chi PT$ applicability range.

Nevertheless, new nonperturbative  schemes imposing unitarity and
still using the chiral Lagrangians have emerged enlarging the
convergence of the chiral expansion. In \cite{ramonet} the inverse 
amplitude method (IAM) is used in one channel and
good results are obtained for the $\sigma$, 
$\rho$ and $K^*$ regions, amongst
others, in $\pi\pi$ and $\pi K$ scattering. 
In \cite{oop, ollerpaco} the method 
is generalized to include coupled channels and one is able to 
describe very well the meson-meson scattering and all the associated
resonances up to about 1.2 GeV. A more general approach is 
used in \cite{oo} by means of the N/D method, in order to include
the exchange of some preexisting resonances explicitly,
which are then responsible for the values
of the parameters of the fourth order chiral Lagrangian.

Here we shall follow the work \cite{oop} since it
provides the most complete study of  
the different meson-meson scattering channels, including the
mesonic resonances and their properties up to 1.2 GeV. In particular, 
this method yields a
resonance in the $I=0,J=1$ channel, the $\omega_8$ 
 resonance, related to the $\phi$, and this allows us  to obtain the 
strong contribution to 
the $\phi\rightarrow\pi\pi$
decay. We also consider electromagnetic
contributions at tree level which turn out to be dominant and were already
considered in \cite{bramon,genz}.

In order to calculate the contribution of an intermediate photon
to the $\phi\rightarrow\pi\pi$ decay, let us consider the effective 
Lagrangian for vector mesons presented in \cite{Ecker}, which
is written in terms of the SU(3) pseudoscalar meson matrix $\Phi$
and the antisymmetric vector tensor field $V_{\mu\nu}$ defined in 
\cite{Ecker}

 \begin{equation}
 {\cal L}_2[V(1^{-\,-})]=\frac{F_V}{2\sqrt{2}}
\langle V_{\mu\nu}f_+^{\mu\nu}\rangle+
\frac{i\,G_V}{\sqrt{2}}\langle V_{\mu\nu}u^\mu u^\nu\rangle,
\label{Lag}
 \end{equation}
where ``$\langle\;\rangle$" indicates the SU(3) trace.
 In order to introduce the physical states $\phi$ and $\omega$,
we assume ideal mixing between the 
 $\omega_1$ and $\omega_8$ vector resonances and hence  
taking into account that the $\omega_1$ does not couple to pairs of mesons at
 the order of eq. (1), the
 coupling of the $\phi$ is easily deduced from that of the $\omega_8$ by simply
 multiplying the results of the $\omega_8$ by the factor $-\frac{2}{\sqrt{6}}$.
With these ingredients and the standard $\gamma \pi \pi$ coupling 
    we can write the contribution of a
 Feynman diagram with the $\phi$ going to a photon which then couples to
 a pair of pions, and which is given by
\begin{equation}
i\,{\cal L}_{\phi\pi^+\pi^-}=i e^2 \frac{\sqrt{2}\,F_V}{3\,M_\phi}
 \epsilon^\mu(\phi) (p_+-p_-)_\mu\,F(M_\phi^2),
\label{Lphipipi}
 \end{equation}
where $p_+$ and $p_-$ are, respectively, 
the momenta of positive and negative pions and
$F(q^2)$ is the pion electromagnetic form factor, 
which at the $\phi$ mass is given by $F(M_\phi^2)=-1.56+i\,0.66$.
  This can be compared with the coupling of the $\phi$ to  $K^+\,K^-$, 
or $K^0\,\bar{K^0}$, which can be obtained from the 
$G_V$ term in Eq.\ (\ref{Lag}) and reads
\begin{equation}
i\,{\cal L}_{\phi K^+ K^-}=-i\, g_{\phi K^+ K^-} \,
 \epsilon^\mu(\phi) \,(p_+-p_-)_\mu\,, \quad
g_{\phi K^+ K^-}=\frac{M_\phi\,G_V}{\sqrt{2}\,f^2}\,,
\label{Lphikk}
\end{equation}
which provides the right $\phi$ decay width with a value of $G_V=54.3$
MeV . 

  By analogy to Eq.\ (\ref{Lphikk}), Eq.\ (\ref{Lphipipi}) 
gives a $\phi$ coupling to $\pi^+\pi^-$
\begin{equation}
g_{\phi \pi^+ \pi^-}^{(\gamma)}
=-\frac{\sqrt{2}}{3}\, e^2 \frac{F_V}{M_\phi}
\,F(M_\phi^2),
\label{g8phikk}
\end{equation}
which  
provides the $\phi\rightarrow \pi^+ \pi^-$ decay width with the
 tree level photon mechanism.  
With a value of $F_V= 154$ MeV from the $\rho\rightarrow e^+e^-$ decay
\cite{Ecker} and using the coupling of Eq.\ (\ref{g8phikk}) one obtains 
a branching ratio to the total $\phi$ width 
of $1.7 \times 10^{-4}$.

In order to evaluate the strong contribution to the process we consider the
 $K\bar{K}\rightarrow\pi^+\pi^-$ amplitude corrected from isospin violation 
 effects due to quark mass differences. The method used is based on 
the chiral unitary approach
to the meson-meson interaction followed in \cite{oop,ollerpaco}. The 
technique starts from the $O(p^2)$ and 
$O(p^4)$ $\chi PT$ Lagrangian and uses the
 IAM in coupled channels, generalizing 
the one channel version of the IAM developed in \cite{ramonet}.

 Within the 
coupled channel formalism, the partial wave amplitude 
is given in the IAM by the matrix equation
\begin{equation}  
T=T_2\,[T_2-T_4]^{-1}\,T_2,
\label{IAM}
\end{equation}  
where $T_2$ and $T_4$ are  $O(p^2)$ and $O(p^4)$ $\chi PT$
partial waves, respectively. In principle $T_4$
would require a full one-loop calculation, but it was shown in
\cite{oop} that it can be very well approximated by
\begin{equation}  
\hbox{Re}\,T_4\simeq T_4^P+T_2\,\hbox{Re}G\,T_2
\label{Ret4}
\end{equation}   
where $T_4^P$ is the tree level polynomial contribution coming 
from the ${\cal L}_4$
chiral Lagrangian and $G$ is a diagonal matrix for the loop function 
of the intermediate two meson propagators which are regularized in  \cite{oop}
 by means of a momentum cut-off.

In the present case, in which  isospin is broken explicitly
and  $J=1$,
we are dealing with three two-meson states:
$K^+ K^-$, $K^0 \bar{K}^0$ and $\pi^+ \pi^-$, that we will call
1, 2 and 3, respectively. The amplitude is a $3\times3$ matrix
whose elements are denoted as $T_{ij}$.
The $T_2$ and $T_4^P$ 
amplitudes used in the present work and calculated 
in the isospin breaking
case, are collected in the appendix of \cite{jaojr}.
  The fit of the
  phase shifts and inelasticities is carried out here
  in the isospin limit, as done in \cite{oop} and there are several sets
  of $L_i$ coefficients which  give rise to equally acceptable fits.

We write in table \ref{tabla2} the values of the coefficients of the
different sets of chiral parameters. 
The  corresponding results for the phase
shifts and inelasticities can be seen in \cite{jaojr} where it is
 shown that the small differences 
in the results appear basically only in the 
$a_0(980)$ and $\kappa(900)$ resonance regions, 
where data have also larger errors or are very scarce.

In order to evaluate the contribution to the $\phi \rightarrow \pi^+ \pi^-$
coupling from the strongly interacting sector we evaluate the 
$K^+K^- \rightarrow K^+ K^-$ amplitude ($T_{11}$) and the 
$K^+ K^- \rightarrow\pi^+ \pi^-$
amplitude ($T_{13}$) near the pole of the $\omega_8$ resonance 
which in our case appears around $M_{\omega_8}=920$ MeV. Close to the 
$\omega_8$ pole the
amplitudes obtained numerically are then driven by the exchange
of an $\omega_8$.

By assuming a coupling of the type of Eq.\ (\ref{Lphikk}) for the $\omega_8$
to $K^+ K^-$ and $\pi^+ \pi^-$, these two amplitudes, close to the $\omega_8$ pole,
 are given by
\begin{eqnarray}
T_{11}&=&g_{\omega_8 K^+ K^-}^2\frac{1}{P^2-M^2_{\omega_8}}\,4\,
\vec{p}_K\cdot\vec{p}_{K'}\nonumber\\
T_{13}&=&g_{\omega_8 K^+ K^-}\,g_{\omega_8 \pi^+ \pi^-}
\frac{1}{P^2-M^2_{\omega_8}}\,4\,
\vec{p}_K\cdot\vec{p}_{\pi}.
\label{t11t13}
\end{eqnarray}
where $\vec{p}_i$ is the three-momentum of the $i$ particle in the CM frame.

By looking at the residues of the amplitudes $T_{11}$, 
$T_{13}$ in the $\omega_8$ pole we can get the products
$g_{\phi K^+K^-}\, g_{\phi K^+K^-}$ and 
$g_{\phi K^+ K^-}\, g_{\phi \pi^+\pi^-}$. Thus, defining 
   \begin{equation}
 \label{Qij}
 Q_{ij}=\lim_{P^2\rightarrow
M^2_{\omega_8}} (P^2-M^2_{\omega_8})\frac{T_{ij}}{4\,\vec{p_i}\cdot\vec{p_j}}
 \end{equation}
   we obtain the ratio of the $g_{\phi K^+K^-}$ to $g_{\phi K^+K^-}$
   by means of the ratio of $Q_{13}$ to $Q_{11}$,
   and hence taking  $g_{\phi K^+K^-}$  from Eq.\ (\ref{Lphikk}), we get 
the value for $g_{\phi \pi^+ \pi^-}^{(s)}$. Then,
 by adding the above contribution with that of
Eq.\ (\ref{g8phikk}) we can   
 obtain the  $\phi\rightarrow\pi^+ \pi^-$
 decay width.  We have
taken $F_V \, G_V > 0$, as demanded by vector meson dominance \cite{Ecker}.

\begin{table}[t]
\begin{scriptsize}
\begin{tabular}{|c|c|c|c|c|c|c|c|c|c|}
\hline
& $\hat L_1$ & $\hat L_2$ & $\hat L_3$ & $\hat L_4$ & $\hat L_5$ & 
$2\hat L_6+\hat  L_8$ & $\hat L_7$ & $q_{max}$ 
& BR$_{\phi\rightarrow\pi\pi}$\\ \hline
set 1&0.91&1.61&-3.65&-0.25&1.07&0.58&-0.4&666 MeV& 1.3$\times 10^{-4}$\\ \hline
set 2&0.91&1.61&-3.65&-0.25&1.07&0.58&0.05&751 MeV& 1.0$\times 10^{-4}$\\ \hline
set 3&0.88&1.54&-3.66&-0.27&1.09&0.68&0.10&673 MeV& 1.3$\times 10^{-4}$\\ \hline\hline
& $L_1$ & $L_2$ & $L_3$ & $L_4$ & $L_5$ & $2L_6+L_8$ & $L_7$ & $\mu$ 
& \\ \cline{1-9}
ChPT&0.4&1.4&-3.5&-0.3&
1.4&0.5&-0.4&$M_\rho$&\rule[.5mm]{1cm}{.2mm}\\
ref.\cite{cual}&$\pm$0.3&$\pm$0.3&$\pm$1.1&$\pm$0.5&$\pm$0.5&$\pm$0.3
&$\pm$0.2&&\\ \hline
\end{tabular}
\end{scriptsize}
\it\caption{\it Different sets of chiral parameters ($\times 10^{-3}$ )
that yield reasonable fits to
the meson-meson scattering phase shifts and the corresponding
$\phi\rightarrow\pi\pi$ branching ratio prediction. We have used a hat to 
differentiate them from those obtained for standard ChPT. However, 
as it is explained in \cite{oop}, we still 
expect them to be relatively similar once the appropriate scales are chosen
(roughly $\mu\simeq1.2\,q_{max}$, see \cite{oop} for details).}
\label{tabla2}
\end{table}

Each set of chiral parameters has then been
used in the isospin-breaking amplitudes given in the appendix of \cite{jaojr}, 
obtaining a
value of  $BR(\phi\rightarrow\pi\pi)$ given in table \ref{tabla2}. The
dispersion of the results provides an estimate of the systematic
theoretical uncertainties.

 From table \ref{tabla2}, we obtain, 
after taking into account the strong contributions
\begin{equation}
  Br(\phi\rightarrow\pi\pi)_{\mbox{\scriptsize tree+strong}}\simeq 
  (1.2\pm 0.2) \times 10^{-4}
\end{equation}
On the other hand,
 explicit calculations of the absorptive part of the
$\eta\gamma$ intermediate channel \cite{bramon} give a contribution of about 
1/4 of the kaon loops.  In order to estimate the uncertainties from neglecting
the photonic loops we take a conservative estimate and consider them of
the same magnitude as the strong interaction correction, and, hence,
add an extra $\pm 0.5\times 10^{-4}$
uncertainty. Adding in quadrature the errors from the different sources, 
our final result is the
band of values:
\begin{equation}
BR(\Phi\rightarrow\pi\pi)\simeq 0.7\;\hbox{to}\; 1.7\times 10^{-4} ,
\end{equation}
which is compatible with
the present PDG average within errors and lies just between the results of the
two recent experiments, which are much more precise, but mutually incompatible.
     
\section{The \boldmath $\phi$ \unboldmath radiative decay into
\boldmath $\pi^0 \pi^0 \gamma$ \unboldmath and 
\boldmath  $\pi^0 \eta \gamma$ \unboldmath }

 The $\phi$ meson cannot decay into two pions or $\pi^0 \eta$ in the isospin 
 limit. The decay into two neutral pions is more strictly forbidden by symmetry
 and the identity of the two pions. As a consequence the decay of the
 $\phi$
 into $\pi^0 \pi^0 \gamma$ and $\pi^0 \eta \gamma$ is forbidden at tree
 level. However, the $\phi$ decays into two kaons and the processes described can
 proceed via the loop diagrams depicted in Fig.\ 1 where the intermediate states
 in the loops stand for $K \bar{K}$.

\begin{figure}[t]
\centerline{
\hbox{\psfig{file=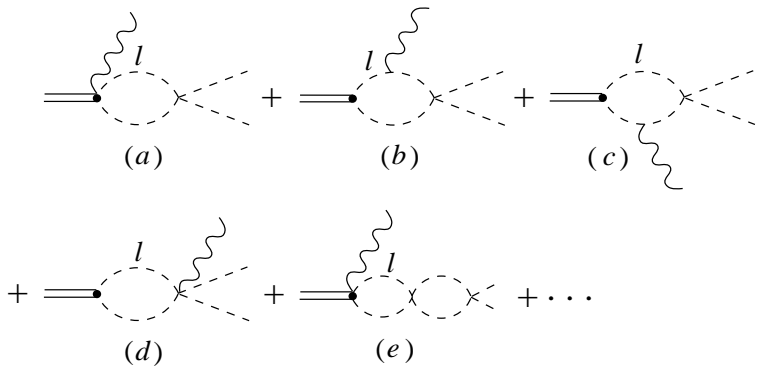,width=10cm,angle=-90}}
}
\caption{\it Diagrams for the decay $\phi \rightarrow \pi^0 \pi^0 \gamma$.}
\label{fig1}
\end{figure}

 The evaluation of the diagrams of Fig.\ 1 is done in \cite{marco}. 
 The terms with $G_V$ of Eq.\  (2) contribute to all the
  diagrams in the figure. However, the $F_V$ term of Eq.\  (2) only contributes
  to  the diagrams containing the contact vertex 
  $\phi\rightarrow \gamma K \bar{K}$, like diagrams (a), (e). The idea
  follows closely the work of \cite{Bra} but for the
  treatment of the final state interaction of the mesons one uses here the
  norperturbative chiral techniques. In this case for L=0, which is the only
  partial wave needed, one can use the results of \cite{OllOse1}, where it is
   proved that the use of the Bethe Salpeter equation in connection  with 
  the lowest order chiral Lagrangian and a suitable cut off in the loops gave 
  a good description of the meson meson scalar sector. Furthermore, in
  \cite{Oller} it was proved that the meson meson amplitude in those diagrams
  factorized on shell. The loops of type (a), (b) and (c) can be summed up
  using arguments of gauge invariance following the  techniques of 
  \cite{Oller,CloIsgKum} and lead to a finite amplitude. On the other hand, the
  terms involving $F_V$ and a remnant momentum dependent term from the $G_V$
  Lagrangian in Eq.\  (2) only appear in the contact vertex 
  $\phi\rightarrow \gamma K \bar{K}$,
  and the diagrams of type (b), (c) are now not present. Hence, in this case the
  only loop function involved is the one of two mesons which is regularized as
  in \cite{OllOse1} for the problem of the meson meson scattering. The average 
  over polarization of the $\phi$ for the modulus square of $t$ matrix is then 
  easily written and for the case of $\pi^0 \pi^0 \gamma$ decay one finds

\begin{eqnarray*}        
\bar{\sum} \sum |t|^2 = \frac{2}{3} e^2 
\left| \frac{M_{\phi} G_V}{f^2\sqrt{3}} \tilde G_{K^+ K^-}
t_{K \bar{K}, \pi \pi}^{I=0} 
+ \frac{K}{f^2\sqrt{3}} \left(\frac{F_V}{2} - G_V\right) G_{K^+ K^-}
t_{K \bar{K}, \pi \pi}^{I=0} \right|^2
\end{eqnarray*}
For the $\phi \rightarrow \pi^0 \eta \gamma$ case we have

\begin{eqnarray*}        
\bar{\sum} \sum |t|^2 = \frac{4}{3} e^2  
\left| \frac{M_{\phi} G_V}{f^2\sqrt{2}}  \tilde G_{K^+ K^-}
t_{K \bar{K}, \pi \eta}^{I=1} 
+ \frac{K}{f^2\sqrt{2}} \left(\frac{F_V}{2} - G_V\right)
 G_{K^+ K^-}
t_{K \bar{K}, \pi \eta}^{I=1}
\right|^2
\end{eqnarray*}
where $\tilde G_{K^+ K^-}$ and $G_{K^+ K^-}$ are the loop functions
mentioned above.

\begin{figure}[t]
\centerline{
\hbox{\psfig{file=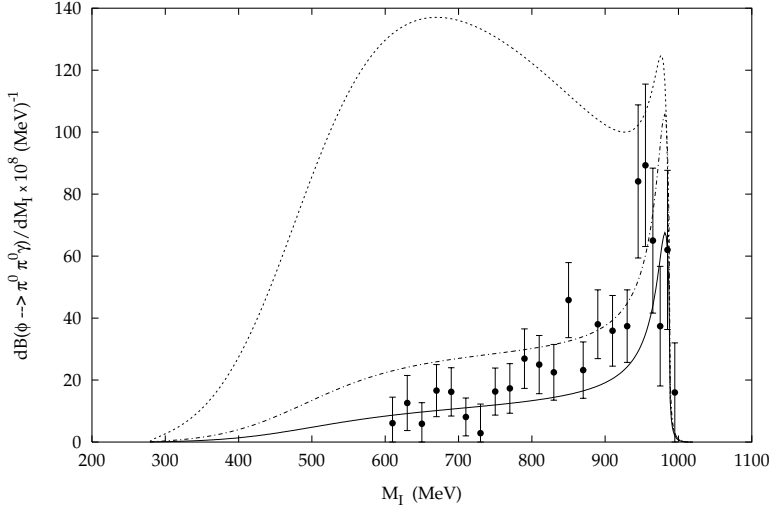,width=10cm,angle=-90}}
}
\caption{\it Distribution $dB /dM_I$ for the decay
$\phi \rightarrow \pi^0 \pi^0 \gamma$, with $M_I$ the invariant mass of
the $\pi^0 \pi^0$ system. Solid line: our prediction, with $F_V G_V >0$.
Dashed line: result taking $F_V G_V <0$.
The data points are from \cite{Novo} and only
statistical errors are shown. The systematic errors are similar to the
statistical ones \cite{Novo}. The intermediate, dot-dashed curve
corresponds to the results obtained using the $G_V$ and $F_V$ parameters
of the $\rho$ decay.}
\label{fig2}
\end{figure}

 We have evaluated the invariant
mass distribution for these decay channels and in Fig.\ 2 we plot
the distribution $dB/dM_I$ for $\phi \rightarrow \pi^0 \pi^0 \gamma$ 
which allows us to see the
$\phi \rightarrow f_0 \gamma$ contribution since the $f_0$ is
the important scalar resonance appearing in the
$K^+ K^- \rightarrow \pi^0 \pi^0$ amplitude \cite{OllOse1}. The results are
obtained using $G_V$=55 MeV and $F_V$=165 MeV, which are suited to describe
the $K \bar{K}$ and $e^+ e^-$ decay of the $\phi$. 
The solid curve shows our prediction, with $F_V G_V >0$, the sign
predicted by vector meson dominance, as we quoted above. The dashed
curve is obtained considering  $F_V G_V <0$. In addition we show also
the results of the intermediate dot-dashed curve which correspond to taking for
$G_V$ and $F_V$ the parameters of the $\rho$ decay, $G_V$=69 MeV and 
$F_V$=154 MeV. We compare
our results with the recent ones of the Novosibirsk experiment \cite{Novo}.
We can see that the shape of the spectrum is relatively well reproduced
considering statistical and systematic errors (the latter ones not shown in
the figure). The results considering $F_V G_V <0$ are
in complete disagreement with the data.

The finite total branching ratio which we find for the 
$\phi \rightarrow \pi^0 \pi^0 \gamma$ decay 
is $0.8 \times 10^{-4}$, which is slightly smaller than
the result given in  \cite{Novo},
$(1.14\pm 0.10\pm 0.12)\times 10^{-4}$, where the first error is statistical
and the second one systematic. The result given in
\cite{CMDpi0pi0} is
$(1.08\pm 0.17\pm 0.09)\times 10^{-4}$, compatible with our prediction.
Should we use the values for $F_V$ and  $G_V$ of the $\rho$ decay we would obtain
$1.7 \times 10^{-4}$.
The branching ratio obtained for the case
$\phi \rightarrow \pi^0 \eta \gamma$ is $0.87 \times 10^{-4}$. The
results obtained at Novosibirsk are \cite{pi0eta} $(0.83 \pm 0.23)\times 10^{-4}$
and \cite{CMDpi0pi0}
$(0.90\pm 0.24\pm 0.10)\times 10^{-4}$. Should we use the values for $F_V$ and  
$G_V$ of the $\rho$ decay we would obtain
$1.6 \times 10^{-4}$.
The spectrum, not shown, is dominated by the $a_0$ contribution.

 The results reported here are two examples of the successful application of the 
 chiral unitary techniques. A recent review of multiple applications of these
 methods can be seen in \cite{oor}.

\vspace{0.1cm}

\section{Acknowledgments}

We would also like to acknowledge 
  financial support from the DGICYT under contracts PB96-0753 and 
  AEN97-1693 and from the EU TMR network Eurodaphne, contract no.
  ERBFMRX-CT98-0169.

\end{document}